\documentclass[preprint]{aastex}
\usepackage[normalem]{ulem}

\shorttitle{Faint Fuzzy Star Clusters in NGC 1023}
\shortauthors{Br\"uns et al.}

\begin{document}

\title{Faint Fuzzy Star Clusters in NGC 1023 as Remnants of Merged Star Cluster Complexes}

\author{R.C. Br\"uns and P. Kroupa}
\affil{Argelander-Institut f\"ur Astronomie, Universit\"at Bonn, Auf dem H\"ugel 71, D-53121 Bonn, Germany}
\email{rcbruens@astro.uni-bonn.de, pavel@astro.uni-bonn.de}

\and

\author{M. Fellhauer}
\affil{Departamento de Astronom\'ia, Universidad de Concepci\'on, Casilla 160-C, Concepci\'on, Chile}
\email{mfellhauer@astro-udec.cl}

\begin{abstract} \label{abstract}
In the lenticular galaxy NGC\,1023 a third population of globular clusters (GCs), called faint fuzzies (FFs), 
was discovered next to the blue and red GC populations by Larsen \& Brodie. While these FFs have colors 
comparable to the red population, the new population is fainter, larger ($R_{\rm eff} > 7$ pc) and, most 
importantly, shows clear signs of co-rotation with the galactic disk of NGC\,1023.
We present N-body simulations verifying the hypothesis that these disk-associated FFs are related to the 
young massive cluster complexes (CCs) observed by Bastian et. al in M51, who discovered a mass-radius relation 
for these CCs. Our models have an initial configuration based on the observations from M51 
and are placed on various orbits in a galactic potential derived for NGC\,1023. All 
computations end up with a stable object containing 10 to 60\% of the initial CC mass after an integration time of 5 Gyr. 
A conversion to visual magnitudes demonstrates that the resulting objects cover exactly the observed range 
for FFs.
Moreover, the simulated objects show projected half-mass radii between 3.6 and 13.4 pc, in good agreement with the 
observed FF sizes.
We conclude that objects like the young massive CCs in M51 are likely progenitors of the 
FFs observed in NGC\,1023.
\end{abstract}


\section{INTRODUCTION} \label{introduction}
A large number of galaxies show a bimodal color distribution of their globular cluster (GC) systems, indicating 
a metal poor blue and a metal rich red GC system \citep[][and references therein]{west04}. \cite{larbro00} studied 
the GC system of the nearby \citep[9.8 Mpc,][]{ciard02} lenticular galaxy NGC\,1023 and discovered a clear bimodal color 
distribution. The average effective radius of the blue GCs in NGC\,1023 is about \mbox{2.0 pc}, while the average radius of 
red compact GCs is slightly smaller ($R_{\rm eff} \sim$ 1.7 pc), as observed also for other galaxies \citep[e.g.][]{lar01}. 
The sizes for the blue clusters are as large as $R_{\rm eff}$ = 7 pc, while the sizes of the red clusters extend up to 
$R_{\rm eff}$ = 18 pc. The objects larger than 7 pc are generally fainter than the red compact objects. 
The spatial distribution of these objects appears to be associated with the lenticular disk of \mbox{NGC\,1023}. Due to their 
faint and extended appearance this new star cluster population was called "Faint Fuzzies" (FFs). 
 
\cite{bro02} performed spectroscopic observations of the GCs in NGC\,1023 and found that the FFs 
show a clear sign of co-rotation with the galactic disk, while the compact red GCs show no sign of rotation, suggesting 
that the FFs are indeed a separate population associated with the disk of NGC\,1023. 
They averaged all FF spectra to achieve a high enough signal-to-noise ratio to estimate the 
age of these objects that appears to be older than 7--8 Gyr. \cite{bur05} further analyzed the 
distribution and radial velocities of the FFs in NGC\,1023 and suggested that they were the remnants of a past 
gravitational interaction forming a ring-like structure.
Further observations also discovered FFs in the lenticular galaxy NGC 3384 \citep{bro02}, the 
dwarf irregular galaxy NGC\,5195 \citep{lee05}, a number of
galaxies of the Virgo cluster \citep{peng}, and the lenticular galaxy NGC 1380 \citep{chisa}. 

Observations have shown that young star clusters are often not isolated, but are part of larger structures called 
cluster complexes (CCs). \cite{bast05} observed massive young star cluster complexes in the disk of the spiral 
galaxy M51 and found a scaling relation between the masses and the radii of the CCs that is comparable to the 
relation of the progenitor giant molecular clouds. This relation constrains the parameter space for the CC models 
in our simulations.

\cite{krou98} studied the dynamical evolution of CCs observed to be forming in the interacting Antennae
galaxies, and \cite{fell02} performed the first N-body simulations indicating that FFs can be formed by merging of star clusters 
in CCs. The work presented in this paper is a follow-up study of these early 
computations. The simulations were performed to test the hypothesis that the FFs are the remnants of the young 
massive CCs as observed by \cite{bast05} in M51. 

In Section \ref{simulations}, we describe the method and the parameters used for the calculations. In Section \ref{results}, 
we present the results of the simulations, which will be discussed in Section \ref{discussion}.

\section{NUMERICAL SIMULATIONS} \label{simulations}
\subsection{Method} \label{method}
The simulations were performed with the particle-mesh code SUPERBOX \citep{fell00}. 
The code solves the Poisson equation on a system of Cartesian grids. The code has a hierarchical grid structure. The 
local universe is covered by a fixed coarse grid which contains the orbit of the CC around the center 
of NGC 1023. The CC orbits in an analytical galactic potential (disk+bulge+halo). In order to get good resolution of the star clusters 
two grids with high and medium resolution are focused on each star cluster following their trajectories. The individual high resolution 
grids cover an entire star cluster, whereas the medium resolution grids of every star cluster embed the whole initial CC. All grids 
contain 64$^{3}$ grid cells. The coordinate system is chosen such that the disk of  \mbox{NGC 1023} lies in the x-y-plane.

\subsection{Basic Parameters} \label{basic_parameters}
\subsubsection{An Estimate of the Gravitational Potential of NGC\,1023} \label{potential_ngc1023}
NGC\,1023 is a disk-dominated SB0-galaxy at a distance of 9.8 Mpc \citep{ciard02} in the constellation Perseus. 
It shows a well behaved rotation in the inner part, while the observed radial velocities in the outer regions 
appear to be more random \citep{noord08}. This behavior is most likely related to a past 
interaction, which transformed the galaxy into an SB0 lenticular galaxy. \cite{noord08} argue that the photometry 
of the disk is consistent with a passive fading since 1 Gyr after the galaxy lost its gas.

\begin{figure}
\epsscale{0.6}
\plotone{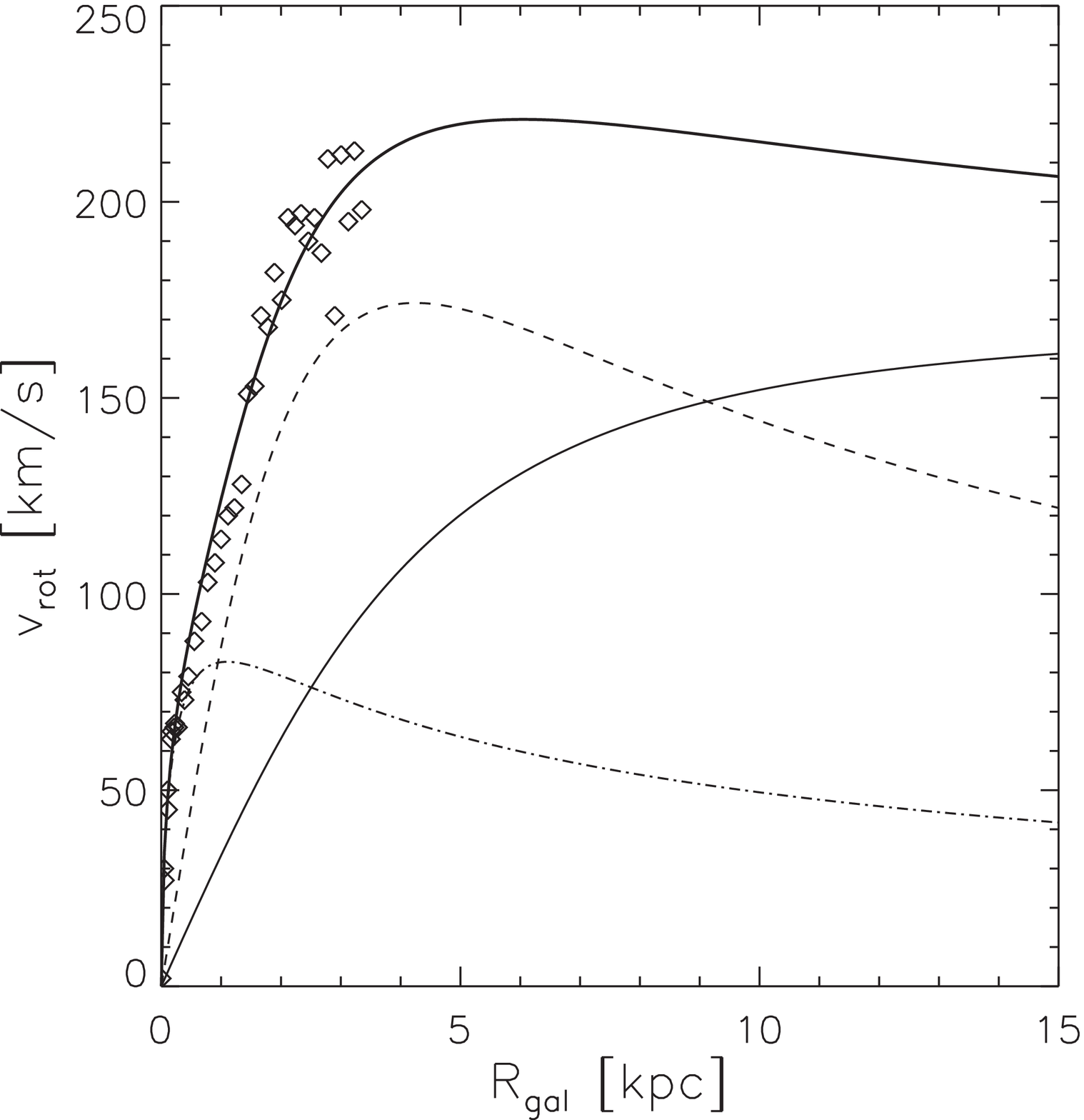}
\caption{Rotation curve of a model of NGC 1023 used for the numerical simulations: the circular velocity is derived from the chosen analytical potential consisting 
 of a Miyamoto-Nagai disk potential (dashed line), a spherical bulge potential (dashed dotted line) and a logarithmic halo (solid line). 
 Summing up the three components yields the total potential. The diamond symbols represent the data available 
 for the inner part (Simien \& Prugniel 1997).\label{fig1}}
\end{figure}

However the exact rotation curve is not required for the simulations because a good guess of the potential is sufficient to study the evolution 
of CCs 5 to 8 Gyr ago. In our simulations NGC 1023 is represented by an analytical potential, which consists of a disk, a bulge, and a 
halo component. For objects moving on circular orbits within the galactic plane, the results of the simulations depend mainly on the total force 
at a given distance, but very weakly on the exact choice of bulge, disk, and halo parameters. We use the observed velocities in the center of 
the galaxy from \cite{simpru97} to constrain the rotation curve in the region of $R_{\rm gal} <$ 4  kpc and assume a flat rotation curve for 
the outer part (Fig. \ref{fig1}). 

The disk is modeled by a Miyamoto-Nagai potential \citep{miya1975}
\begin{eqnarray}
\Phi_{disk}(R,z)= -\frac{G\,M_{d}}{\sqrt{R^{2}+(a_{d}+\sqrt{z^{2}+b_{d}^{2}})^{2}}}, 
\end{eqnarray}
with $M_{\rm d}=5.5 \cdot 10^{10}$ M$_{\odot}$, $a_{\rm d}=2.8$ kpc, and $b_{\rm d}=0.2$ kpc. \\
The bulge is represented by a Hernquist potential \citep{hern1990}
\begin{eqnarray}
\Phi_{bulge}=-\frac{G \, M_{b}}{r+a_{b}},
\end{eqnarray}
with $M_{\rm b}=0.7 \cdot 10^{10}$ M$_{\odot}$, and $a_{\rm b}=1.1$ kpc.
The halo is a logarithmic potential
\begin{eqnarray}
\Phi_{halo}(r)=\frac{1}{2} \, v_{0}^{2} \, {\rm ln}(r^{2}+r_{halo}^{2}), 
\end{eqnarray}
with $v_{\rm 0}=170$ km s$^{-1}$, and $r_{\rm halo}=5.0$ kpc.\\

For this project, we neglected heating sources from non-axisymmetric components like Giant Molecular Clouds (GMC)
and spiral arms. This is a good approximation for compact objects with masses of about 10$^{5}$ M$_{\odot}$. 
\cite{gieles06} demonstrated that star clusters with such masses cannot be destroyed by individual
encounters with GMCs, if the extended nature of GMCs is taken into account. Close encounters between GMCs and massive 
star clusters act as tidal heating on the clusters. The corresponding disruption time is much larger than the integration 
time of the simulations presented in this paper. \cite{gieles07} analyzed the effect of spiral arms passages on the evolution
of star clusters and found for massive clusters as considered in this paper a disruption time of the order 100 Gyr.

\subsubsection{Initial Configuration of the Cluster Complexes} \label{initial_configuration}
\cite{bast05} derived a mass-radius relation for the CCs in M51 that constrains the parameter 
space for our CC models. They found 11 CCs with ages younger than 10 Myr in the disk of M51 that have 
sizes between $85 - 240$ pc, and cover a mass range of 0.3 - 3 $\cdot$ 10$^{5}$ M$_{\odot}$.
Most of these CCs show a massive concentration of clusters 
potentially merging in their centers and a couple of isolated clusters in their vicinity \citep[see fig. 2 and 3 in][]{bast05}. 
The size of a CC is defined as the point where the measured color is equal to that of the background stars in the disk. 
This radius corresponds to the cutoff radius of our models. 
The spatial distribution of the CCs is connected to the spiral arms of the disk. Most complexes are found at the 
outer edges of the spiral arms. The CCs show a mass-radius relation comparable to the one of the progenitor 
giant molecular clouds. 

\begin{figure}
\epsscale{0.6}
\plotone{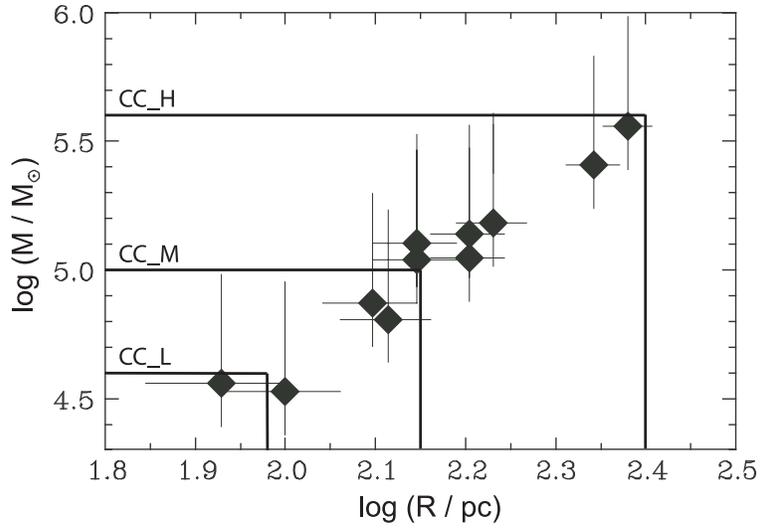}
\caption{The mass versus radius relation for CCs in M51 (Bastian et al. 2005). The 11 complexes, indicated by diamonds, 
have sizes between $\sim 85$ and $\sim 240$ pc and cover a mass range of $0.3-3 \cdot 10^{5}$ M$_{\odot}$. 
The size of the complex is defined as the radius where the CC is still 
clearly distinguishable from the background light of the galaxy. It is the cutoff radius for our models, 
which cover the high- and low-mass end of the Bastian relation (CC\_H and CC\_L) and 
a point in-between (CC\_M).\label{fig2}}
\end{figure}

We choose the parameters of CCs in the simulations such that the models cover the two extremes (CC\_H 
and CC\_L) and a point in-between (CC\_M) of the Bastian relation (Fig. \ref{fig2}).

The star clusters building up the CCs in our simulations are Plummer spheres \citep{plum1911, krou08}. The density distribution 
of the spheres is truncated at \mbox{$R_{\rm cut} = 5 R_{\rm pl}$}. Each star cluster consists of $N_{\rm 0}^{\rm SC}$ = 100\,000 particles. 
The CC models are diced according to a Plummer distribution with masses and radii taken from the Bastian relation 
(Fig. \ref{fig2}). Our models resemble the CCs observed by \cite{bast05} with a high density concentration in their 
centers.

\cite{bast05} estimated a very high star formation efficiency within the CCs, using the CO-to-H$_{2}$ conversion 
factor for M51 from \cite{bos02} and a Salpeter IMF \citep{salpeter55}. Due to the high star formation efficiency, gas expulsion has only little 
influence on the evolution of the merging CCs \citep{fell05}. Consequently, the distorting effect of 
gas expulsion was not considered in the simulations.

\section{RESULTS} \label{results}
We have carried out 25 different numerical simulations to study the influence of varying initial CC conditions and 
orbital parameters. The computations are presented in the following way. In Section \ref{different_cc_masses} we investigate 
the evolution of CCs in three mass regimes (CC\_H, CC\_M, CC\_L) of the Bastian relation. In Section \ref{big_sim} we consider a maximal 
model with 80 star clusters. For comparison, computations without an external gravitational field are presented in 
Section \ref{sim_wtf}. 

\subsection{Configurations Covering the Cluster Complex Mass Range} \label{different_cc_masses}

We investigate the evolution of a CC configuration in three different mass regimes (Fig.~\ref{fig2}) for circular orbits at 
galactic distances of $R_{\rm gal}$ = 2, 3, 5, 8, 12 kpc (Section \ref{circular_orbit}) and an eccentric orbit (Section \ref{eccentric_orbit}). 
The number of star clusters comprising the CC is fixed to $N_{\rm 0}^{\rm CC}$ = 20 in all models. In order to get comparable starting conditions 
for the models CC\_H, CC\_M, and CC\_L we introduce a parameter 
\begin{eqnarray} \alpha = \frac{R_{pl}^{SC}}{R_{pl}^{CC}}, \end{eqnarray} 
where $R_{\rm pl}^{\rm SC}$ and $R_{\rm pl}^{\rm CC}$ are the Plummer radius of a single star cluster and the Plummer radius of the CC, respectively
 \citep{fell02a}. The parameter $\alpha$ is a measure of how densely the CC is filled with star clusters for an equal number 
 $N_{\rm 0}^{\rm CC}$ of star clusters. In general high values of $\alpha$ accelerate the merging process because the star clusters already 
 overlap in the center of the CC, whereas low values hamper the merging process. We choose a medium value of $\alpha$ = 0.08, 
 which leads to a reasonable range of the Plummer radii of the individual star clusters. The grids were also scaled to get the same 
 relative resolution for all models. The chosen initial parameters are displayed in Table \ref{tbl-1}.

\subsubsection{Simulations on Circular Orbits} \label{circular_orbit}

We have performed 15 simulations on circular orbits holding the initial star 
cluster configuration and the number of star clusters in the complex fixed. The CC masses were restricted to the high- and low-mass end 
(CC\_H\_20 and CC\_L\_20) of the Bastian relation and a value in-between (CC\_M\_20). The orbital parameters are given in \mbox{Table \ref{tbl-2}}.
The circular velocities, $v_{\rm circ}$, at the different galactic distances, $R_{\rm gal}$, were derived from the rotation curve of NGC 1023 
(Fig. \ref{fig1}). The orbital periods lie between 70 and 350 Myr. 

The timescale of the merging process is very short. The merger object forms within a few crossing times of the CC. 
In all cases the merging leads to a stable final object. The properties of the merger objects at $t$ = 5 Gyr are displayed in 
Table \ref{tbl-3}. The number of merged star clusters varies from 10 to 16. The number increases with galactic distance as the 
influence of the tidal field becomes weaker. 

\begin{figure}
\epsscale{0.83}
\plotone{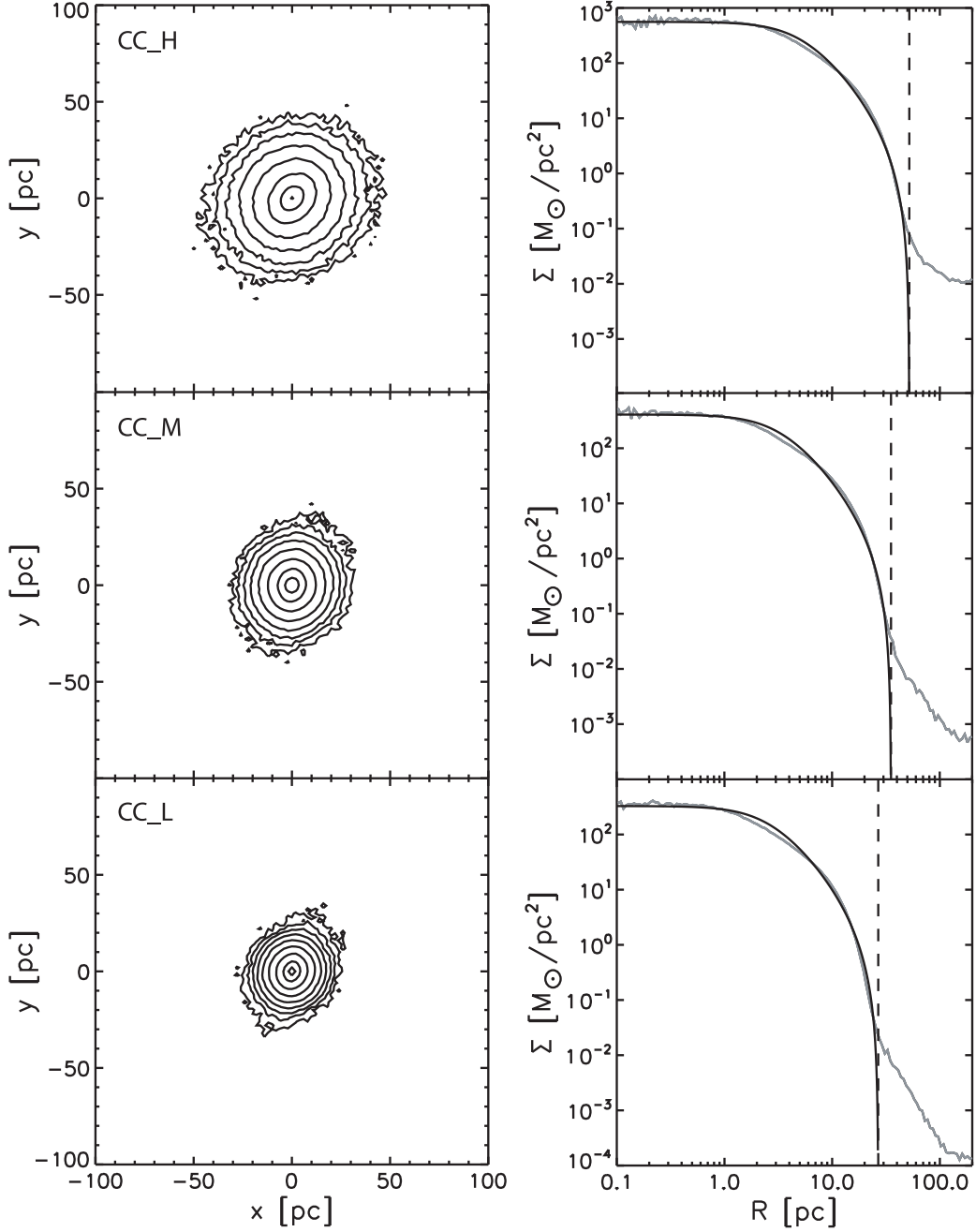}
\caption{Left: Contour plots at $t$ = 5 Gyr on the x-y-plane for the computations CC\_H\_5\_20, CC\_M\_5\_20 and CC\_L\_5\_20 at a galactic 
 distance of $R_{\rm gal} = 5$ kpc. 
 The lowest contour level corresponds to 5 particles per pixel. The pixel size is 4 pc$^{2}$. 
 This yields 0.25 M$_{\odot}$ pc$^{-2}$ (CC\_H), 0.0625 M$_{\odot}$ pc$^{-2}$ (CC\_M) and 0.025 M$_{\odot}$ pc$^{-2}$ (CC\_L). 
 The contour levels increase further by a factor of 3. Right: Surface density profile corresponding to the contour plots 
 of CC\_H\_5\_20, CC\_M\_5\_20 and CC\_L\_5\_20. The profiles are fitted by King models. The dashed vertical lines denote the 
 tidal radii.\label{fig3}}
\end{figure}

In all cases we calculate surface density profiles and fit these by a King profile \citep{kin66}. Three exemplary final 
merger objects after 5 Gyr are shown in Fig. \ref{fig3} 
as contour plots on the x-y-plane and as corresponding surface density profiles. 

The tidal radii vary from $R_{\rm t} = 18.8$ pc for simulation CC\_L\_2\_20 at $R_{\rm gal} = 2$ kpc to \mbox{$R_{\rm t}$ = 88.9 pc} for 
CC\_H\_12\_20 at $R_{\rm gal} = 12$ kpc.
The enclosed mass of the merger objects is defined as the mass within the tidal radius. After 5 Gyr 20 to 60 \% of the initial CC mass
is bound to the merger object. The half-mass radius is the radius
of the sphere, where half of the mass is enclosed. However, as observers can only derive a projected half-mass radius, we calculate also
a projected value defined as the projected radius in the surface density plot within which half of the mass is included. The projected
half-mass radius is slightly smaller than the three-dimensional half-mass radius (Table \ref{tbl-3}) and corresponds to the observed effective
radius, $R_{\rm eff}$. 

The enclosed masses and half-mass radii are increasing with galactic distance $R_{\rm gal}$, leading to small compact merger objects at 
$R_{\rm gal} = 2$ kpc and more extended massive ones at $R_{\rm gal} = 12$ kpc. Comparing the different models CC\_H\_20, CC\_M\_20, and CC\_L\_20 
the tidal field has the largest impact on model CC\_H\_20 impeding 
the merging process. An estimate of the influence of the tidal field is given by the parameter 
\begin{eqnarray} \beta =\frac{R_{cut}^{CC}}{R_{t}^{CC}} \end{eqnarray} \citep{fell02a}, which is the fraction of the cutoff radius $R_{\rm cut}^{\rm CC}$ of the CC and 
its tidal radius $R_{\rm t}^{\rm CC}$. The larger the value of $\beta$, the larger the impact. The tidal radius is estimated by using
the Jacobi limit for a satellite on a circular orbit \citep{bintre1987},
\begin{eqnarray}
\begin{tabular}{ll}
  $R_{t}^{CC}$& =  $\left( \frac{M_{CC}}{3 \cdot M_{gal}(R_{gal})} \right)^{\frac{1}{3}} \cdot R_{gal}$ \\
              & $\approx \left( \frac{M_{CC} \cdot G}{3 \cdot v_{c}^{2}(R_{gal})} \right)^{\frac{1}{3}} \cdot R_{gal}^{\frac{2}{3}}$. \\
\end{tabular}
\end{eqnarray}
The values for the initial tidal radii of the CCs and the corresponding parameter $\beta$ of the individual models are shown in Table \ref{tbl-2}.
The $\beta$-values are decreasing from model CC\_H\_20 to CC\_L\_20 and from small galactic distances of $R_{\rm gal} = 2$ kpc to large 
distances of $R_{\rm gal} = 12$ kpc. Therefore model CC\_H\_20 suffers 
a higher percental mass-loss than simulations CC\_M\_20 and CC\_L\_20 and for each simulation the impact of the tidal force on the CC 
is larger for small galactic distances $R_{\rm gal}$.

\subsubsection{Cluster Complexes on an Eccentric Orbit} \label{eccentric_orbit}
We performed 3 simulations on eccentric orbits to test the  
influence of the orbit on the properties of the final merger object. The orbit was 
chosen such that the distance of the CC to the center of NGC 1023 varies between 3 kpc and 8 kpc. 
The orbital period is about 100 Myr. Again the computations were performed for the three different mass regimes (Fig. \ref{fig2}).
Table \ref{tbl-3} shows the properties of the final merger objects.

For simulation CC\_H\_ecc\_20 the value for the enclosed mass lies between the
values of the circular orbits at $R_{\rm gal} = 2$ kpc and $R_{\rm gal} = 3$ kpc. The half-mass and tidal radii agree more with those obtained for
the 2 kpc circular orbit. The properties of the final object of simulation CC\_M\_ecc\_20 and CC\_L\_ecc\_20 are similar to those of the circular
3 kpc orbit. Thus, for the eccentric orbit the peri-galactic distance determines the overall properties of the final merger object.\\  

\subsubsection{Different Initial Configurations} \label{different_initial_configs}
For comparison we also performed simulations for two additional initial configurations in the medium mass regime 
(CC\_M\_5\_c2\_20 and CC\_M\_5\_c3\_20). 
Different seeds were used for the random number generator to generate two different initial configurations having the same 
input parameters as CC\_M\_20 (Table \ref{tbl-1}). The initial configurations vary considerable due to the low number of 
clusters constituting the CC.  
The models were placed at a distance of $R_{\rm gal}$ = 5 kpc on a circular orbit. 

The two configurations also lead to a fast merger within the first few 
crossing times. After 5 Gyr the final objects have masses of $2.5 \cdot 10^{4}$ M$_{\odot}$ (CC\_M\_5\_c2\_20) and  $2.9 \cdot 10^{4}$ M$_{\odot}$ (CC\_M\_5\_c3\_20). 
The corresponding projected half-mass radii are 5.5 pc and 5.6 pc (Table \ref{tbl-3}). In the case of configuration CC\_M\_5\_c2\_20 16 star clusters 
have merged. Whereas for configuration CC\_M\_5\_c3\_20 the number of merged star clusters is only 13 although its final mass is larger.
The final mass of the merger object does not scale with the number of merged star clusters as the newly formed object looses  
considerable mass on its orbit around the center of NGC 1023. 
The final parameters of the merger objects are slightly smaller than those of the corresponding model CC\_M\_5\_20 (see Sect. \ref{circular_orbit}).
While the enclosed mass and the half-mass radius of a merger object depend considerably on the initial configuration, the general order of
magnitude stays the same.

\subsection{A Model With a Large Number of Star Clusters}  \label{big_sim}

In this section we consider a maximal CC with $N_{\rm 0}^{\rm CC}$ = 80 star clusters. The parameters for the computation
are presented in Table \ref{tbl-1}. 
The general parameters for the CCs are taken from model CC\_H\_20. Due to the larger number of star clusters, 
the individual star clusters have lower masses, which are comparable to the star cluster masses of simulation CC\_M\_20. 
The complex is placed only on one circular orbit at a 
distance of $R_{\rm gal}$ = 5 kpc, because this computation is very time-consuming.

\begin{figure}
\epsscale{0.6}
\plotone{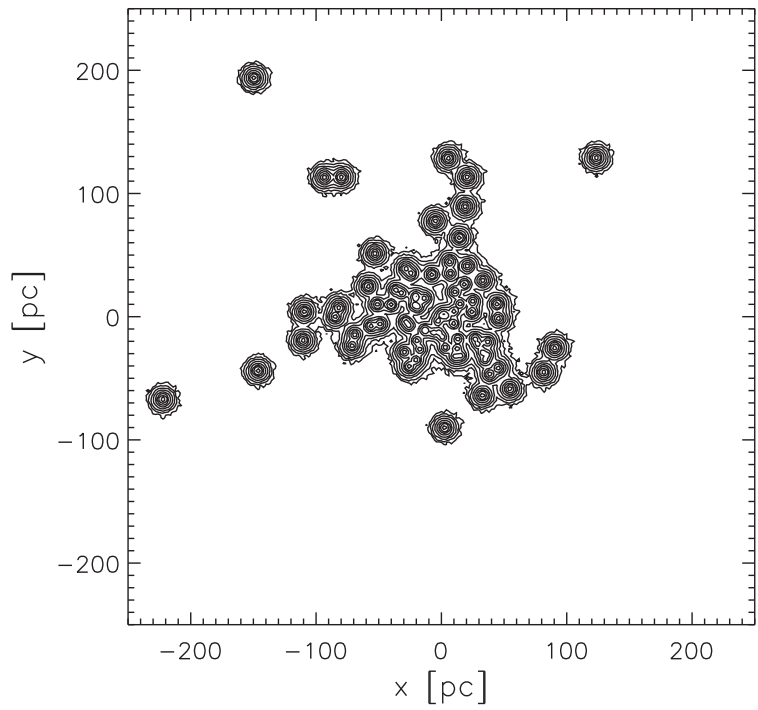}
\caption{Contour plot on the x-y-plane of the initial distribution of the big computation CC\_H\_5\_80 with 80 star clusters.
 The lowest contour level corresponds to 5 particles per pixel. The pixel size is 4 pc$^{2}$. The other contours 
 increase by a factor of three.\label{fig4}}
\end{figure}

\begin{figure}
\epsscale{0.6}
\plotone{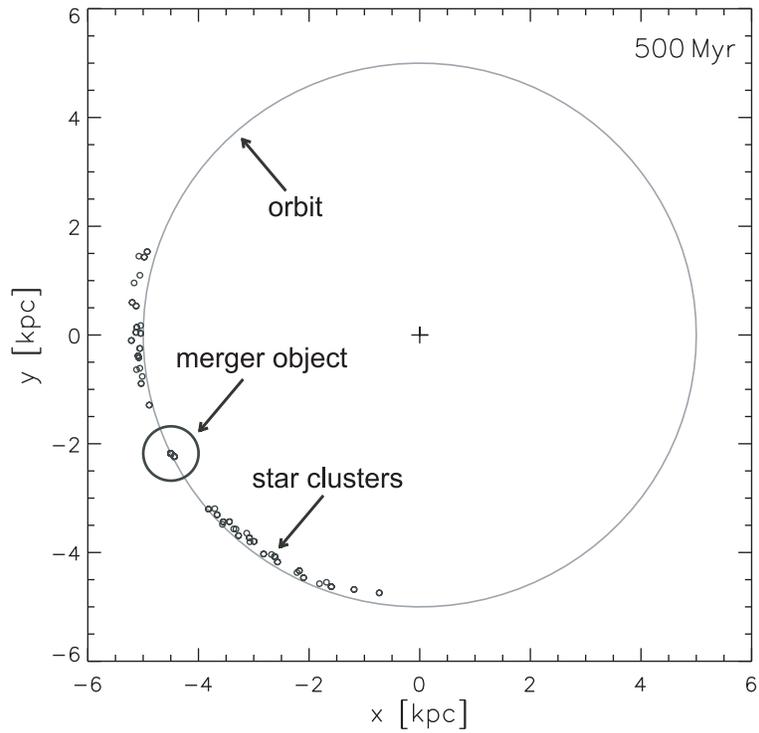}
\caption{Illustration of the spatial distribution of the star clusters of computation CC\_H\_5\_80 after 500 Myr. The CC moves on a circular 
orbit around the center of NGC 1023 (cross). It consists of a large merger object and a leading 
and trailing arm of about 50 unmerged star clusters. The circles indicate the positions but not the sizes
of the star clusters. \label{fig5}}
\end{figure}

\begin{figure}
\epsscale{0.65}
\plotone{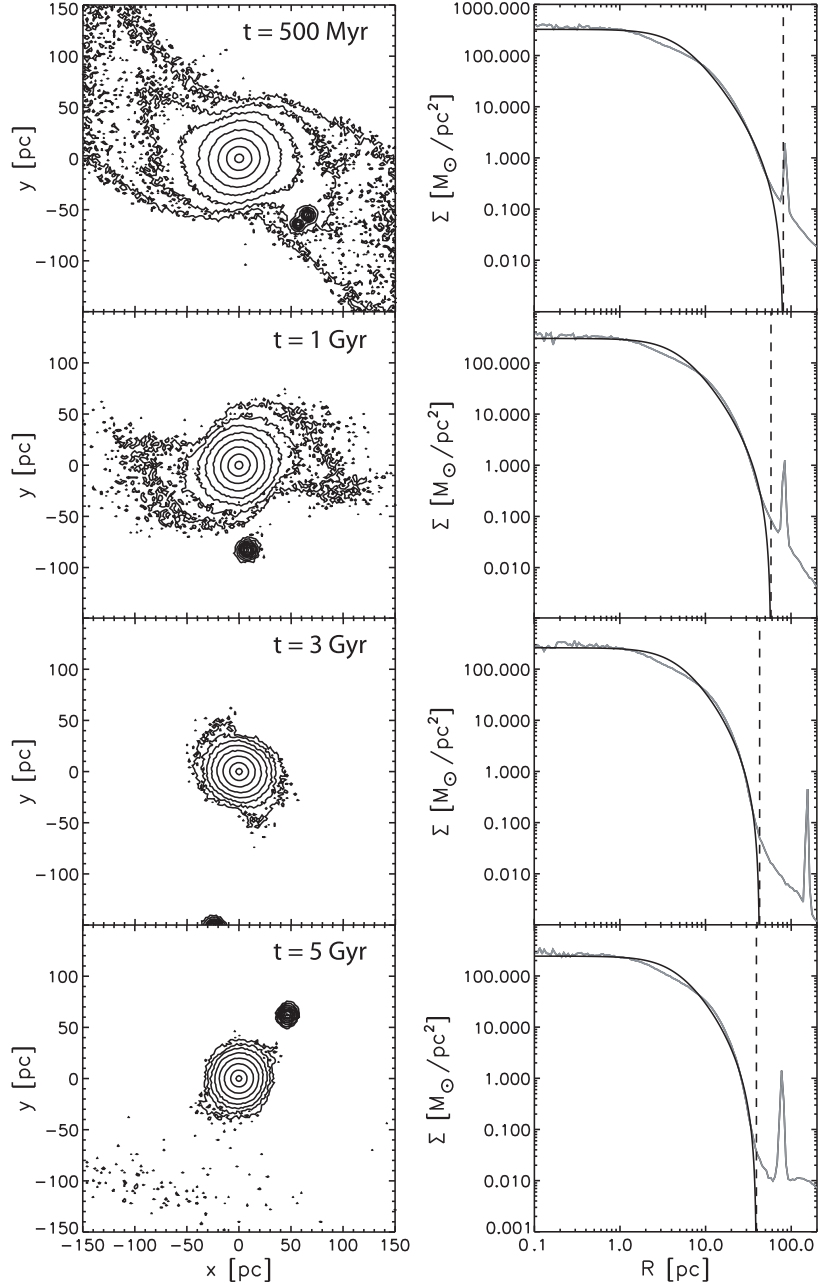}
\caption{Time evolution of the merger object in simulation CC\_H\_5\_80 at a galactic distance of $R_{\rm gal} = 5$ kpc. 
 Left: Contour plots on the x-y-plane displayed at $t$ = 0.5, 1, 3, 5 Gyr. 
 The lowest contour level corresponds to 5 particles per pixel. The pixel size is $4$ pc$^{2}$ 
 which corresponds to 0.0625 M$_{\odot}$ pc$^{-2}$. 
 The contour levels increase further by a factor of 3. Right: Surface density profiles corresponding to the contour plots.
 The profiles are fitted by King models. The dashed vertical lines denote the tidal radii. \label{fig6}}
\end{figure}

The initial star cluster distribution is displayed in Fig. \ref{fig4} as a contour plot on the x-y-plane. The distribution
is quite extended and has a concentration of star clusters in its center as observed for CCs in M51 by 
Bastian et al. (2005). Within the first 100 Myr a massive merger object forms and the unmerged star clusters align along the orbit 
(Fig. \ref{fig5}). There is still a close companion star cluster as a satellite of the merger object. The time evolution of the merger 
object is shown in Fig. \ref{fig6} as contour plots. The snapshots were taken at $t$ = 0.5, 1, 3 and 5 Gyr. At $t$ = 500 Myr (top panel) 
the merger object has already formed. It is very extended and shows prominent tidal arms. The major mass loss occurs within the first Gyr.
Later the merger object becomes more compact and its tidal arms are less pronounced. After 5 Gyr the merger object has reached an 
almost stable state. The mass loss is small and the tidal radius hardly changes any more. 

Table \ref{tbl-3} lists the properties of the final merger object which resulted from the merger of 32 star clusters. 
Its mass is only 9 \% of the initial CC mass and its projected half-mass radius is 8.5 pc. 

\subsection{Cluster Complexes without External Gravitational Field} \label{sim_wtf}
For comparison we performed simulations without an external gravitational potential (i.e. $R_{\rm t} = \infty$) for the three 
models CC\_H\_20, CC\_M\_20, CC\_L\_20, and the maximal configuration with 80 star clusters, CC\_H\_80. 
In case of the three computations with varying CC masses 17 out of 20 star clusters merge within the first 500 Myr. 
After 5 Gyr nearly all star clusters have merged and formed an object with projected half-mass radii of 22.5 pc (CC\_H\_inf\_20), 
12.7 (CC\_M\_inf\_20) and 9.0 (CC\_L\_inf\_20) which contain more than 90 \% of the initial CC mass (Table \ref{tbl-3}). 

For the simulation with 80 star clusters (CC\_H\_inf\_80) the merging process is considerably slower. 
In the first 500 Myr 46 star clusters merge. The newly formed merger object is surrounded by several unmerged star 
clusters which are found preferentially in its close vicinity. During the next 500 Myr another 23 star clusters merge 
into the merger object. After 5 Gyr 79 star clusters have merged resulting in a final object with a projected half-mass radius
of 45.6 pc and a mass of $3.9 \cdot 10^{5}$ M$_{\odot}$ corresponding to 97.5 \% of the initial CC mass. 

The individual star clusters in CC\_H\_80 have basically the same relative velocities as in CC\_H\_20, but four times lower masses.
This reduces the probability of merging considerably. The resulting longer merging timescale provides the gravitational 
field more time to tear clusters off the complex. This explains the lower number of merged star clusters and 
the relatively low mass in the case of the corresponding model within a tidal field. 

\section{DISCUSSION AND CONCLUSION} \label{discussion}

\begin{figure}
\epsscale{1.0}
\plotone{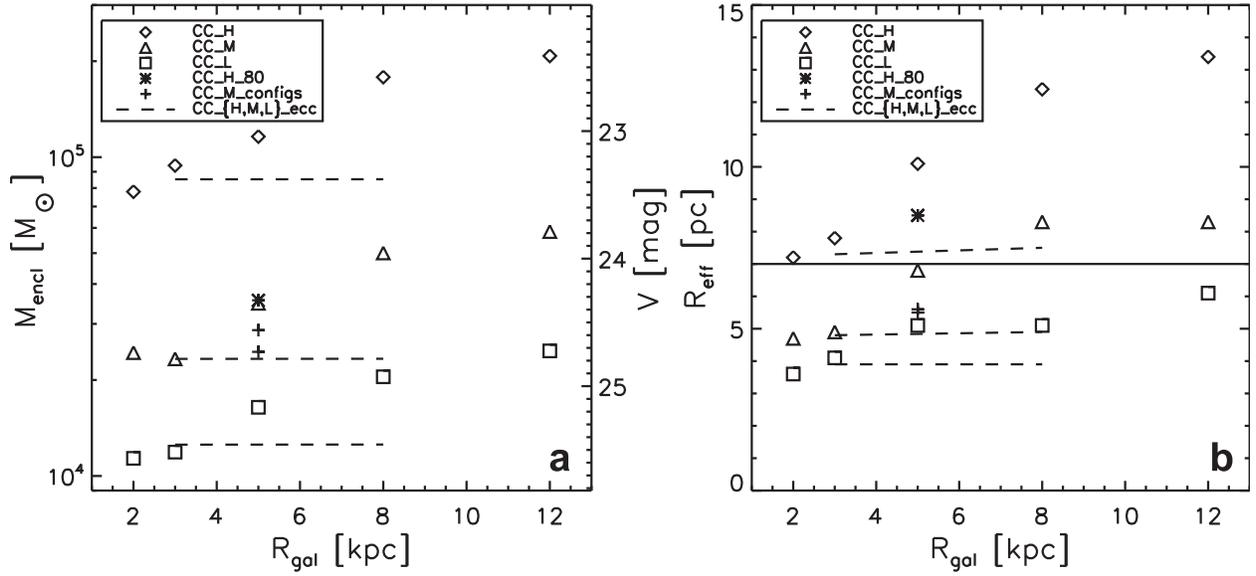}
\caption{{\bf a}: Enclosed mass $M_{\rm encl}$ versus galactic distance $R_{\rm gal}$ for all computations. 
Simulation CC\_H\_20, CC\_M\_20 and CC\_L\_20 were performed for different distances of $R_{\rm gal}$ = 2, 3, 5, 8 and 12 kpc whereas the big 
computation CC\_H\_5\_80 with 80 clusters and the two additional configurations of CC\_H\_5\_20 were only run 
at a distance of $R_{\rm gal} = 5$ kpc. The eccentric orbit models cover a radius range 
of 3 - 8 kpc and are therefore plotted as dashed lines. 
{\bf b}: Effective radius $R_{\rm eff}$ against galactic distance $R_{\rm gal}$ for all simulations as in the left diagram. The horizontal line 
at 7 pc indicates the lower size limit for FFs defined by \cite{larbro00}. \label{fig7}}
\end{figure}

We summarize the major properties for all computations in Fig. \ref{fig7}. The models CC\_H\_20, CC\_M\_20 and CC\_L\_20 were placed on circular orbits 
at 5 different distances whereas the big configuration CC\_H\_5\_80 with 80 star clusters and the two additional configurations were only 
run on a circular orbit at $R_{\rm gal} = 5$ kpc. The 3 computations on an eccentric orbit are indicated by 
dashed lines ranging from 3 to 8 kpc. We plotted the enclosed mass and the projected half-mass radius, 
$R_{\rm eff}$, against the galactic distance from the center of NGC\,1023. 

The enclosed masses were converted into $V$-magnitudes to allow for direct comparison with the observed data, using the formula
\begin{equation}
V = M_{V,solar}-2.5 \cdot \log(M_{\rm FF}/x)+29.97,
\end{equation}
where $M_{\rm V,solar}$ = 4.83 mag is the absolute solar $V$ magnitude, $M_{\rm FF}$ the mass of the FF, $x$ the mass-to-light ratio and
the value of 29.97 the distance modulus derived from the planetary nebula luminosity function \citep{ciard02}. We use a mass-to-light ratio of 
$x$ = 2.3, as observed by \cite{prymy93} for 56 Globular Clusters in the Milky Way. No observational constraints on the M/L-ratio
of FFs are available.

Fig. \ref{fig7}a shows the enclosed mass as well as the corresponding V-magnitudes versus the galactic distance $R_{\rm gal}$. 
The V-magnitudes of the merger objects cover a range from 22.4 mag to 25.6 mag. \cite{larbro00} detected FFs in
the luminosity range 21.4 mag to 25 mag\footnote{\cite{larbro00} did not consider objects below 25 mag. Their approximate \mbox{50 \%} completeness 
limit was 24.5 mag.}, with a median of 23.6 mag, but only one FF is brighter than 22.5 mag. The luminosities of the simulated merger
objects are therefore in very good agreement with the observed luminosities of FFs.  

Figure \ref{fig7}b shows the effective radius of the merged CCs versus galactic distance. The effective radii for all 
simulations range between 3.6 and 13.4 pc. The effective radius criterion of $R_{\rm eff} \ge$ 7 pc from \cite{larbro00} is added 
as a horizontal line. The merger objects of model CC\_H\_20 (circular and eccentric) and the big simulation CC\_H\_5\_80 with 80 star 
clusters lie clearly above the radius limit whereas the computations of model CC\_M\_20 exceed the limit only at galactic distances of 
$R_{\rm gal} \ge$ 8 kpc. 
The lowest mass merger objects have effective radii below 7 pc. However, the effective radius criterion of $R_{\rm eff} \ge$ 7 pc is not 
to be taken as a definite limit but more as a reference value, because below $R_{\rm eff}$ = 7 pc it is not possible to distinguish 
FFs from compact red GCs. 80 \% of the FFs observed by \cite{larbro00} have effective radii below 14 pc. The median of the observed 
effective radii is 10.7 pc. Our results are therefore in good agreement with observations.

The simulations presented in this paper demonstrate that CCs from the high-mass end of the Bastian relation evolve into FF-like 
objects. Our formation scenario suggests that FFs are the remnants of merged CCs formed 
in spiral arms. NGC\,1023 has a prominent bar that extents up to about 3 kpc \citep{debatt02}. In barred spiral
galaxies, spiral arms usually start where the bar ends \citep[e.g.][]{elmegreen}. The absence of FFs in the inner 3 kpc of
NGC\,1023 is therefore consistent with the formation scenario presented in this paper. The Milky Way, which is also a barred galaxy 
with a comparable size to NGC\,1023, has most of its giant molecular clouds in the disk between Galactic radii 
3 to 8 kpc, with a peak between 4 and 5 kpc \citep{dame}. The mass-radius relation found for young massive CCs in M51 is 
comparable to the corresponding relation for giant molecular clouds in M51 \citep{bast05}. Consequently, the radial distribution 
of FFs should in general follow the radial distribution of giant molecular clouds, but modified by the results of Fig. \ref{fig7},
i.e. lower masses and smaller effective radii for decreasing galactic distances. 
In NGC\,1023 FFs were found between projected galactic radii 3 kpc $\le R_{\rm gal}$ $\le$ 10 kpc, with a clear peak at a 
projected radius of 4.5 kpc. The observed radial distribution of FFs is therefore in good agreement with the expected 
distribution from our model. 
 
In summary, the simulated merger objects based on the observed mass-radius relation from \cite{bast05} resemble the observed 
parameters mass/luminosity, size and spatial distribution of FFs in NGC\,1023 very well. The merging of young massive  
CCs is therefore a likely formation scenario for FFs. 

The measured velocities from \cite{bro02} confirm the general co-rotation of the FFs with the disk of NGC\,1023.
\cite{bur05} argued that the annular distribution of FFs on the sky and some deviations from the rotation curve of NGC\,1023 
indicate that the FFs form a ring-like structure with a mean radius of 5 kpc potentially formed during a past tidal encounter. In their
model, FFs are on highly eccentric orbits spanning the entire range between 3 and 8 kpc. 
In our model, FFs are associated with the former spiral arms of NGC\,1023, as the CCs in M51 are associated with the spiral arms 
of the disk. FFs should therefore follow the rotation of the disk of NGC\,1023, which was probably disturbed during a past interaction that transformed
NGC 1023 into a lenticular galaxy.
However, due to the very low luminosities, the observational uncertainties of the velocities are relatively large. \cite{bur05}
re-observed two FFs from \cite{bro02}. The new measurements of these clusters yield 539\,$\pm$\,21 and 
676\,$\pm$\,13 km\,s$^{-1}$, while the previous measurements found 514\,$\pm$\,8 and 725\,$\pm$\,17 km\,s$^{-1}$, respectively. These 
differences are too large for definite statements about deviations from circular orbits.

\cite{bro02} averaged all FF spectra shifted to zero velocity to achieve a high enough signal-to-noise ratio to estimate the 
age of these objects that appears to be older than 7--8 Gyr. They used the same integrated spectrum to derive a mean metellicity
of [Fe/H] = --0.58$\pm$0.24 and a mean alpha-to-iron ratio [$\alpha$/Fe] between +0.3 and +0.6 compared to solar values. 
Enhanced alpha-to-iron ratios are usually associated with rapid, burst-like star formation on short time scales. 
Indeed, the 5 to 8 Myr old CCs observed by \cite{bast05} show a low age spread, indicating that the stellar mass of 
the order of $10^{5}$ M$_{\odot}$ has been formed in a burst-like event.

The FFs need to be further analyzed with respect to additional and more precise kinematical data to allow for a reliable estimate of 
their orbits. In addition, age estimates, metallicities, and alpha-to-iron ratios are needed for individual FFs. The scatter and the 
variation of these parameters with galactic distance will shed light on the FF evolution and further test whether the discussed formation 
scenario is valid.

\acknowledgments
The work of this paper was supported by DFG Grant KR\,1635/14-1. We thank S{\o}ren Larsen for helpful discussions and access to their FF-data.

\clearpage

\clearpage

\begin{deluxetable}{lrrrrrrrrr}
\tablecolumns{10}
\tablewidth{0pc}
\tablecaption{Initial Cluster Complex and Star Cluster Parameters\label{tbl-1}}
\tablehead{
\colhead{}    &  \multicolumn{4}{c}{Cluster Complex} &   \colhead{}   &
\multicolumn{4}{c}{Star Cluster} \\
\cline{2-5} \cline{7-10} 
\colhead{Model} & \colhead{$N_{\rm 0}^{\rm CC\,}$\tablenotemark{a}} & \colhead{$M_{\rm CC\,}$\tablenotemark{b}} & 
\colhead{$R_{\rm cut}^{\rm CC\,}$\tablenotemark{c}} & \colhead{$R_{\rm pl}^{\rm CC\,}$\tablenotemark{d}} &
\colhead{}    & \colhead{$N_{\rm 0}^{\rm SC\,}$\tablenotemark{e}}  & \colhead{$M_{\rm SC\,}$\tablenotemark{f}} & 
\colhead{$R_{\rm cut}^{\rm SC\,}$\tablenotemark{g}}    & \colhead{$R_{\rm pl}^{\rm SC\,}$\tablenotemark{h}} \\
\colhead{} & \colhead{} & \colhead{(M$_{\odot}$)} & \colhead{(pc)} & \colhead{(pc)} &
\colhead{}    & \colhead{}  & \colhead{(M$_{\odot}$)} & \colhead{(pc)}    & \colhead{(pc)} }
\startdata
CC\_H\_20 & 20 & $4 \cdot 10^{5}$ & 250 & 50 & & $100\,000$ & $2 \cdot 10^{4}$ & 20.0 & 4.0 \\
CC\_H\_80 & 80 & $4 \cdot 10^{5}$ & 250 & 50 & & $100\,000$ & $5 \cdot 10^{3}$ & 10.0 & 2.0 \\
CC\_M\_20 & 20 & $1 \cdot 10^{5}$ & 141.5 & 28.3 & & $100\,000$ & $5 \cdot 10^{3}$ & 11.5 & 2.3 \\
CC\_L\_20 & 20 & $4 \cdot 10^{4}$ & 95.5 & 19.1 & & $100\,000$ & $2 \cdot 10^{3}$ & 7.5 & 1.5 \\
\enddata
\tablenotetext{a}{Number of star clusters comprising the CC.}
\tablenotetext{b}{Initial CC mass.}
\tablenotetext{c}{Cutoff radius of the CC.}
\tablenotetext{d}{Plummer radius of the CC.}
\tablenotetext{e}{Number of star cluster particles.}
\tablenotetext{f}{Initial mass of star cluster.}
\tablenotetext{g}{Cutoff radius of star cluster.}
\tablenotetext{h}{Plummer radius of star cluster.}
\end{deluxetable}

\begin{deluxetable}{rrrrrrrrrrrr}
\tablecolumns{12}
\tablewidth{0pc}
\tablecaption{Orbital Parameters\label{tbl-2}}
\tablehead{
\multicolumn{3}{c}{} & \colhead{} & \multicolumn{2}{c}{CC\_H\_20} & \colhead{} & \multicolumn{2}{c}{CC\_M\_20} & \colhead{} & \multicolumn{2}{c}{CC\_L\_20}\\
\cline{5-6} \cline{8-9} \cline{11-12}
\colhead{$R_{\rm gal\,}$\tablenotemark{a}} & \colhead{$v_{\rm circ\,}$\tablenotemark{b}} & \colhead{$T_{\rm orb\,}$\tablenotemark{c}} &\colhead{} & 
\colhead{$R_{\rm t}^{\rm CC\,}$\tablenotemark{d}} &
\colhead{$\beta\,$\tablenotemark{e}} & \colhead{} & \colhead{$R_{\rm t}^{\rm CC\,}$}  & \colhead{$\beta\,$} & \colhead{}&
\colhead{$R_{\rm t}^{\rm CC\,}$} & \colhead{$\beta\,$}  \\
\colhead{(kpc)} & \colhead{(km s$^{-1}$)} & \colhead{(Myr)} & \colhead{} & \colhead{(pc)} &
\colhead{} & \colhead{} & \colhead{(pc)}  & \colhead{} & \colhead{} & \colhead{(pc)} & \colhead{} }
\startdata
2 & 174 & 70.4 & & 42.1 & 5.9 & & 26.5 & 5.3 & & 19.6 & 4.9\\
3 & 202 & 91.1 & & 50.0 & 5.0 & & 31.5 & 4.5 & & 23.2 & 4.1\\
5 & 220 & 139.7 & & 66.5 & 3.8 & & 41.9 & 3.4 & & 30.9 & 3.1\\
8 & 219 & 224.3 & & 91.3 & 2.7 & & 57.5 & 2.5 & & 42.4 & 2.3\\
12 & 212 & 348.5 & & 122.4 & 2.0 & & 77.1 & 1.8 & & 56.8 & 1.7\\
\enddata
\tablenotetext{a}{Distance to the center of galaxy NGC\,1023.}
\tablenotetext{b}{Circular velocity obtained from the rotation curve (Fig. \ref{fig1}).}
\tablenotetext{c}{Orbital period.}
\tablenotetext{d}{Tidal radius of CC at $t$ = 0 (Eq. 6).}
\tablenotetext{e}{$\beta$-parameter at $t$ = 0 (Eq. 5).}
\end{deluxetable}

\begin{deluxetable}{lrrrrrrrr}
\tablecolumns{11}
\tablewidth{0pc}
\tablecaption{Results of the 25 Computations after 5 Gyr\label{tbl-3}}
\tablehead{
\colhead{Model} & \colhead{$R_{\rm gal\,}$\tablenotemark{a}} & \colhead{$N_{\rm M\,}$\tablenotemark{b}} & 
\colhead{$M_{\rm encl\,}$\tablenotemark{c}} & \colhead{$M_{\rm encl\,}$} &
\colhead{$V\,$\tablenotemark{d}} & \colhead{$R_{\rm h\,}$\tablenotemark{e}} & \colhead{$R_{\rm eff\,}$\tablenotemark{f}} & 
\colhead{$R_{\rm t\,}$\tablenotemark{g}} \\
\colhead{} & \colhead{(kpc)} & \colhead{} & \colhead{(M$_{\odot}$)} & \colhead{(\%)} &
\colhead{(mag)} & \colhead{(pc)}  & \colhead{(pc)}  & \colhead{(pc)}}
\startdata
CC\_H\_2\_20 & 2 & 10 & 77923.6 & 19.5 & 23.5 & 8.6 & 7.2 & 37.6\\
CC\_H\_3\_20 & 3 & 12 & 94180.2 & 23.5 & 23.3 & 9.5 & 7.8 & 41.7\\
CC\_H\_5\_20 & 5 & 12 & 116021.0 & 29.0 & 23.0 & 12.3 & 10.1 & 52.0\\
CC\_H\_8\_20 & 8 & 14 & 178244.0 & 44.6 & 22.6 & 15.3 & 12.4 & 64.3\\
CC\_H\_12\_20 & 12 & 15 & 207737.0 & 51.9 & 22.4 & 17.1 & 13.4 & 88.9\\
CC\_H\_ecc\_20 & 3-8 & 12 & 85120.6 & 21.3 & 23.4 & 9.3 & 7.4 & 38.2\\
CC\_H\_inf\_20 & $\infty$ & 19 & 362804.0 & 90.7 & 21.8 & 29.6 & 22.5 & - \\
CC\_H\_5\_80 & 5 & 32 & 35565.1 & 8.9 & 24.3 & 10.6 & 8.5 & 39.2\\
CC\_H\_inf\_80 & $\infty$ & 79 & 390052.0 & 97.5 & 21.7 & 60.8 & 45.6 & - \\
CC\_M\_2\_20 & 2 & 12 & 24366.5 & 24.4 & 24.7 & 5.7 & 4.7 & 24.4\\
CC\_M\_3\_20 & 3 & 12 & 23297.8 & 23.3 & 24.8 & 6.0 & 4.9 & 26.4\\
CC\_M\_5\_20 & 5 & 13 & 34793.1 & 34.8 & 24.4 & 8.3 & 6.8 & 35.3\\
CC\_M\_8\_20 & 8 & 15 & 50045.4 & 50.0 & 24.0 & 10.0 & 8.3 & 43.5\\
CC\_M\_12\_20 & 12 & 16 & 58413.3 & 58.4 & 23.8 & 10.5 & 8.3 & 58.6\\
CC\_M\_ecc\_20 & 3-8 & 12 & 23198.6 & 23.2 & 24.8 & 6.2 & 4.9 & 26.1\\
CC\_M\_5\_c2\_20 & 5 & 16 & 24507.4 & 24.5 & 24.7 & 6.9 & 5.5 & 29.7\\
CC\_M\_5\_c3\_20 & 5 & 13 & 28668.2 & 28.7 & 24.6 & 6.9 & 5.6 & 30.4\\
CC\_M\_inf\_20 & $\infty$ & 19 & 90740.5 & 90.7 & 23.3 & 16.6 & 12.7 & - \\
CC\_L\_2\_20 & 2 & 13 & 11360.3 & 28.4 & 25.6 & 4.4 & 3.6 & 18.8\\
CC\_L\_3\_20 & 3 & 13 & 11879.8 & 29.7 & 25.5 & 4.9 & 4.1 & 20.9\\
CC\_L\_5\_20 & 5 & 14 & 16422.8 & 41.1 & 25.2 & 6.3 & 5.1 & 25.8\\
CC\_L\_8\_20 & 8 & 14 & 20470.2 & 51.2 & 24.9 & 6.5 & 5.1 & 30.3\\
CC\_L\_12\_20 & 12 & 16 & 24679.0 & 61.7 & 24.7 & 7.7 & 6.1& 43.9\\
CC\_L\_ecc\_20 & 3-8 & 13 & 12491.0 & 31.2 & 25.5 & 4.8 & 3.9 & 21.1\\
CC\_L\_inf\_20 & $\infty$ & 20 & 37643.0 & 94.1 & 24.3 & 11.7 & 9.0 & - \\
\enddata
\tablenotetext{a}{Distance to the center of galaxy NGC\,1023.}
\tablenotetext{b}{Number of merged star clusters.}
\tablenotetext{c}{Enclosed mass of merger object.}
\tablenotetext{d}{V-magnitudes (Eq. 7).}
\tablenotetext{e}{Half-mass radius of merger object.}
\tablenotetext{f}{Effective radius, i.e. the projected half-mass radius of the merger object.}
\tablenotetext{g}{Tidal radius of merger object obtained from King fit.}
\end{deluxetable}

\end{document}